\documentclass[onecolumn]{aa} 

\usepackage{graphicx}
\usepackage{color}
\usepackage{subfigure}
\usepackage{longtable}

\usepackage{txfonts}
\begin{document}

   \title{Structural analysis of star-forming blue early-type galaxies}
   

 \subtitle{Merger-driven star formation in elliptical galaxies}

   \author{Koshy George
          \inst{}}

   \institute{Indian Institute of Astrophysics, 2nd Block, Koramangala, 
Bangalore - 560034, India\\
              \email{koshy@iiap.res.in}}


   \date{Received September 12, 2016; accepted December 5, 2016}

 
 \abstract
   {Star-forming blue early-type galaxies at low redshift can give insight to
 the stellar mass growth of L$*$ elliptical galaxies in the local Universe.}
   {We wish to understand the reason for star formation in these otherwise passively evolving red and dead stellar systems. The fuel for star formation can be acquired 
   through recent accretion events such as mergers or flyby.  The signatures of such events
 should be evident from a structural analysis of the galaxy image.}
   {We carried out structural analysis on SDSS $r$-band imaging data of 55 star-forming blue elliptical galaxies, derived
 the structural parameters, analysed the residuals from best-fit to surface brightness distribution, and constructed the galaxy scaling relations.}
   {We found that star-forming blue early-type galaxies are
 bulge-dominated systems with axial ratio $>$ 0.5 and  surface brightness profiles fitted
 by Sersic profiles with index ($n$) mostly $>$ 2. Twenty-three galaxies are found to have
 $n$ $<$ 2; these could be hosting a disc component. The
 residual images of the 32 galaxy surface brightness profile fits show structural features indicative of recent interactions. The star-forming blue elliptical galaxies follow the Kormendy
 relation and show the characteristics of normal elliptical galaxies as far as structural analysis is concerned. There is
a general trend for high-luminosity galaxies to display interaction signatures and high star
formation rates.}
   {The star-forming population of blue early-type galaxies at low redshifts could be 
normal ellipticals that might have undergone a recent gas-rich minor merger event. The star formation in these
 galaxies will shut down once the recently acquired fuel is consumed, following which
 the galaxy will evolve to a normal early-type galaxy.}


\maketitle
%

\section{Introduction}

Early-type galaxies (ETGs) in the local Universe are morphologically ellipticals or lenticulars with little or no ongoing star formation, host evolved stellar population,
 and populate the red sequence on the galaxy colour-magnitude diagram \citep{Faber_1973,Visvanathan_77,Baldry_2004}. Early-type galaxies  populate the massive
 end of the galaxy stellar mass function at least from redshift ($z$) $\sim$ 1 and  constitute a significant fraction of the stellar mass in the local Universe.  Recent observations
 suggest that the stellar mass of field L* ETGs almost doubled from $z$ $\sim$ 1 to 0 \citep{Bell_2004,Faber_2007,Brown_2007}. The observed increase of stellar
 mass in L* ETGs  between $z$ $\sim$ 1 to 0 is in accordance with the current paradigm of $\Lambda$CDM cosmology, which predicts
 hierarchical formation and evolution for galaxies \citep{Delucia_2006}.\\

Galaxies follow different evolutionary paths over the last 8 billion years to account for the observed growth in stellar mass. The galaxies can be at different stages of the evolutionary path,
 which at different redshifts and varying environments can be used to test our current understanding of galaxy formation and evolution (\citealt{Faber_2007} for a detailed description of the various galaxy evolutionary paths).
Star-forming blue spiral galaxies in the absence of continuous supply of molecular gas can quench star formation and migrate
 from the star-forming blue cloud to the passively
 evolving red sequence on the galaxy colour-magnitude diagram. Major mergers between two star-forming spiral galaxies  go through a morphological destructive phase with
 intense star formation that depletes the gas reservoir and makes the merger remnant an ETG on the red sequence \citep{Toomre_77}. Major mergers are more
 common at high redshift when the Universe was less than half of the current age \citep{Conselice2003}. Minor mergers, on the other hand, 
are more frequent and can contribute to the continuous build-up of galaxies in the local Universe \citep{Kaviraj_2009,Kaviraj_2014}. The 
slow build-up of the outer galaxy until the present epoch, as observed in the size evolution of elliptical galaxies, is now believed to be due to minor mergers \citep{Trujillo_2011}.
 Minor mergers can also bring gas and induce star formation in an otherwise gas-poor elliptical galaxy. HI surveys of relaxed nearby elliptical galaxies demonstrate
 the presence of significant amounts of neutral hydrogen, the
origin of which is attributed to external sources like minor mergers \citep{Morganti2006, Oosterloo2010,Serra2014}. 
The processes responsible for bringing  gas to elliptical galaxies can be mergers or gas accretion from the intergalactic medium. Star formation can 
occur from this gas and can account for some of the stellar mass build-up in ETGs. Galaxies can then move from the red sequence to the blue cloud on the galaxy colour-magnitude diagram.\\

The report of a sample of blue star-forming ETGs is interesting in this context (\citealt{schawinski09}, hereafter S09). The elliptical
 morphology of these L$*$ galaxies is  based on visual classifications from the Galaxy Zoo (GZ). They are  found in low-density environments and also
  make up 5.7 $\pm$ 0.4 $\%$ of the low-redshift early-type galaxy population. Blue ellipticals have an  $r$-band 
absolute magnitude (M$r$) $<$ -20.5 and $u-r$ colour $<$ 2.5 mag based on the positions on the line diagnostic diagram \citep{Baldwin_1981}. S09 have classified  
25 $\%$ as only star-forming, 25 $\%$ as both star-forming and active galactic nucleus (AGN), 12 $\%$ as AGN and 38 $\%$ as having no strong emission lines to classify. 
The star formation rates (SFR) of galaxies are estimated from the H$\alpha$ line luminosity using the calibration of 
\citep{Kennicutt_1998} (see S09 for a comparison of SFR estimated with other proxies).  The 
55 purely star-forming blue ETGs (out of a total of 204 blue ETGs) have high star 
formation rates (0.45 to 21 M$\odot$/yr) and very blue colours for any such systems found in the local Universe. Few of these systems
 are reported as ongoing or recently interacting galaxies \citep{George15}. The 55 star-forming blue ETGs are used in this study to
 understand the existence of such systems in the current paradigm of galaxy formation and evolution.  We address the following points: we confirm the elliptical nature, the triggering mechanism 
of star formation in an otherwise red and dead elliptical galaxy, the availability of fuel for star formation, and the fate of blue ETGs.  Structural
 analysis can give insights into the underlying morphology of these objects, and 
the estimation of structural parameters allows us to make galaxy
 scaling relations. We first establish the elliptical nature of blue ETGs by a detailed structural analysis and place these systems in the context of galaxy scaling relations. We adopt a
 flat Universe cosmology with $H_{\rm{o}} = 71\,\mathrm{km\,s^{-1}\,Mpc^{-1}}$, $\Omega_{\rm{M}} = 0.27$, $\Omega_{\Lambda} = 0.73$ \citep{Komatsu_2011}.\\

\section{Data}

The SDSS DR7 $r$-band images of 55 star-forming blue ETGs from S09 were used for the structural analysis. We used the structural parameters data from 
the catalogue of \citet{meert_2015}, which is  constructed by applying two-dimensional point-spread function (PSF) corrected surface brightness profile fits of $7 {\times} 10^{5}$ 
spectroscopically selected galaxies from SDSS DR7.  The 
fitting routine GALFIT \citep{Peng_2002} and analysis pipeline PyMorph \citep{Vikram_2010} were used to estimate the 
structural parameters. The choice of fitting the model (single or double) components to the galaxies to measure the total magnitude and half-light radius was extensively tested based on simulations of mock galaxies by 
\citet{meert_2013}. The authors concluded that a Sersic $+$ exponential component can optimally measure the galaxy global parameters across many galaxy types.
Hence we used the structural parameters estimated with surface brightness profile fitting with a double
 component, Sersic \citep{Sersic1968}, and an exponential profile  to perform a best fit on  the galaxy surface brightness distribution.
  The measured total magnitude, 
effective radius, Sersic index, and 
the ratio between the major and the minor axis (axial ratio $b/a$) of the galaxy brightness distribution were used.  We further analysed the residual data from the 
profile fitting results for a detailed structural 
study (residual images from a private communication with Alan Meert). We also used the structural parameters estimated with 
a surface brightness profile fitting using a single Sersic profile to derive the galaxy global parameters
 and check the dependence of the choice on the number of components that affect our residual analysis. \\

\section{Structural analysis}

\subsection{Star-forming blue early-type galaxies}

\begin{figure*}
  \centering
  \mbox{
\subfigure[\label{subfigure label}]{\includegraphics[width=10.00cm,height=10.00cm,keepaspectratio=true,angle=0]{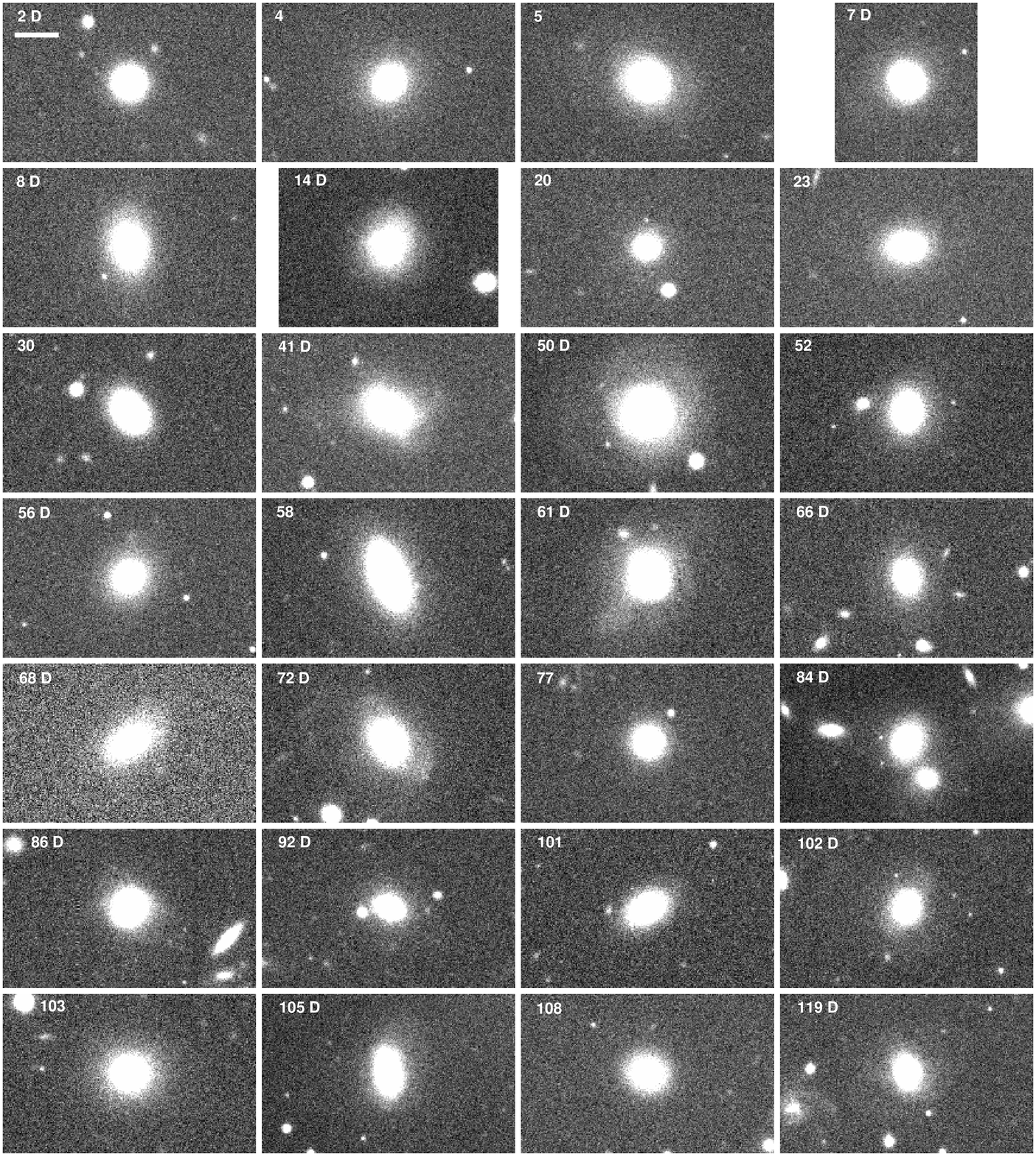}}\quad
\subfigure[\label{subfigure label}]{\includegraphics[width=10.00cm,height=10.00cm,keepaspectratio=true,angle=0]{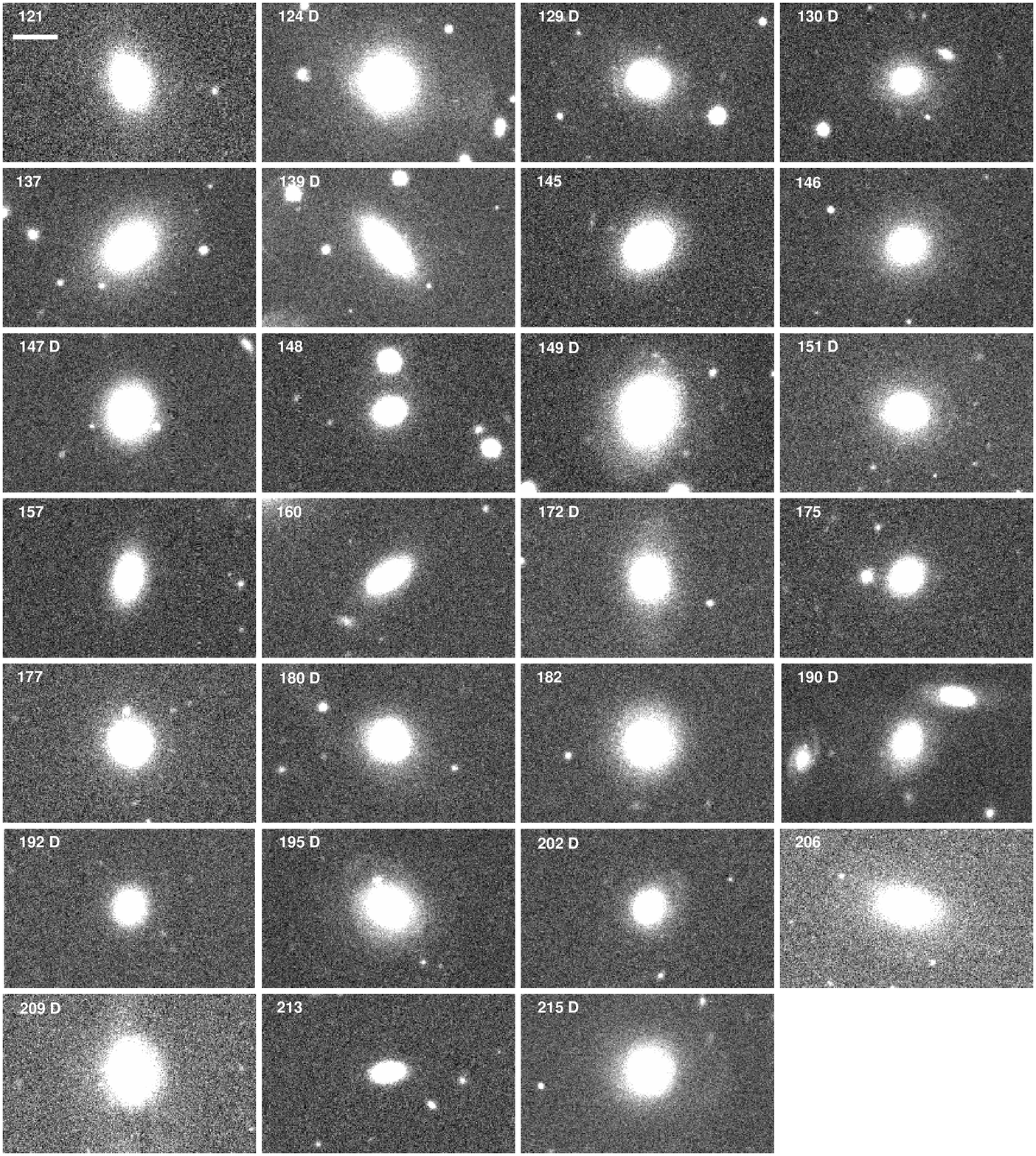}}\quad}
  \caption{SDSS $r$-band images of 55 star-forming blue ETGs taken with 54s integration time. Note the faint features at the galaxy outskirts with
 shallow SDSS imaging. The S09 object identification number is given at the top of each image. The images showing the presence of recent interaction features in residual analysis 
 are marked with D. All images are of the same dimension, and a line of length 15$\arcsec$ is shown in the first image; this corresponds to the typical size of the galaxy.}
  \label{fig:fig1}
\end{figure*}

\begin{figure*}
  \centering
  \mbox{
\subfigure[\label{subfigure label}]{\includegraphics[width=10.0cm,height=10.0cm,keepaspectratio=true,angle=0]{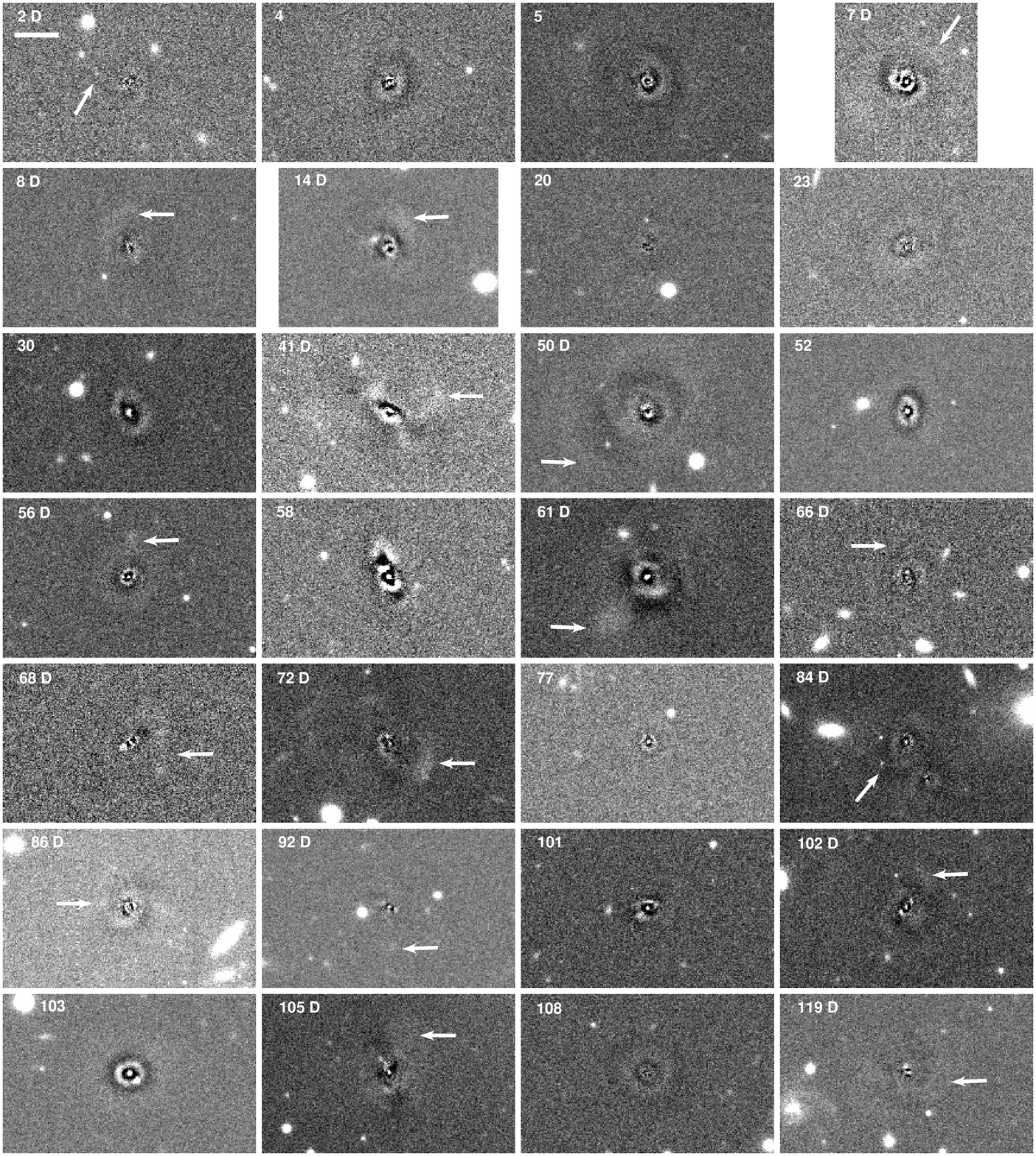}}\quad
\subfigure[\label{subfigure label}]{\includegraphics[width=10.0cm,height=10.00cm,keepaspectratio=true,angle=0]{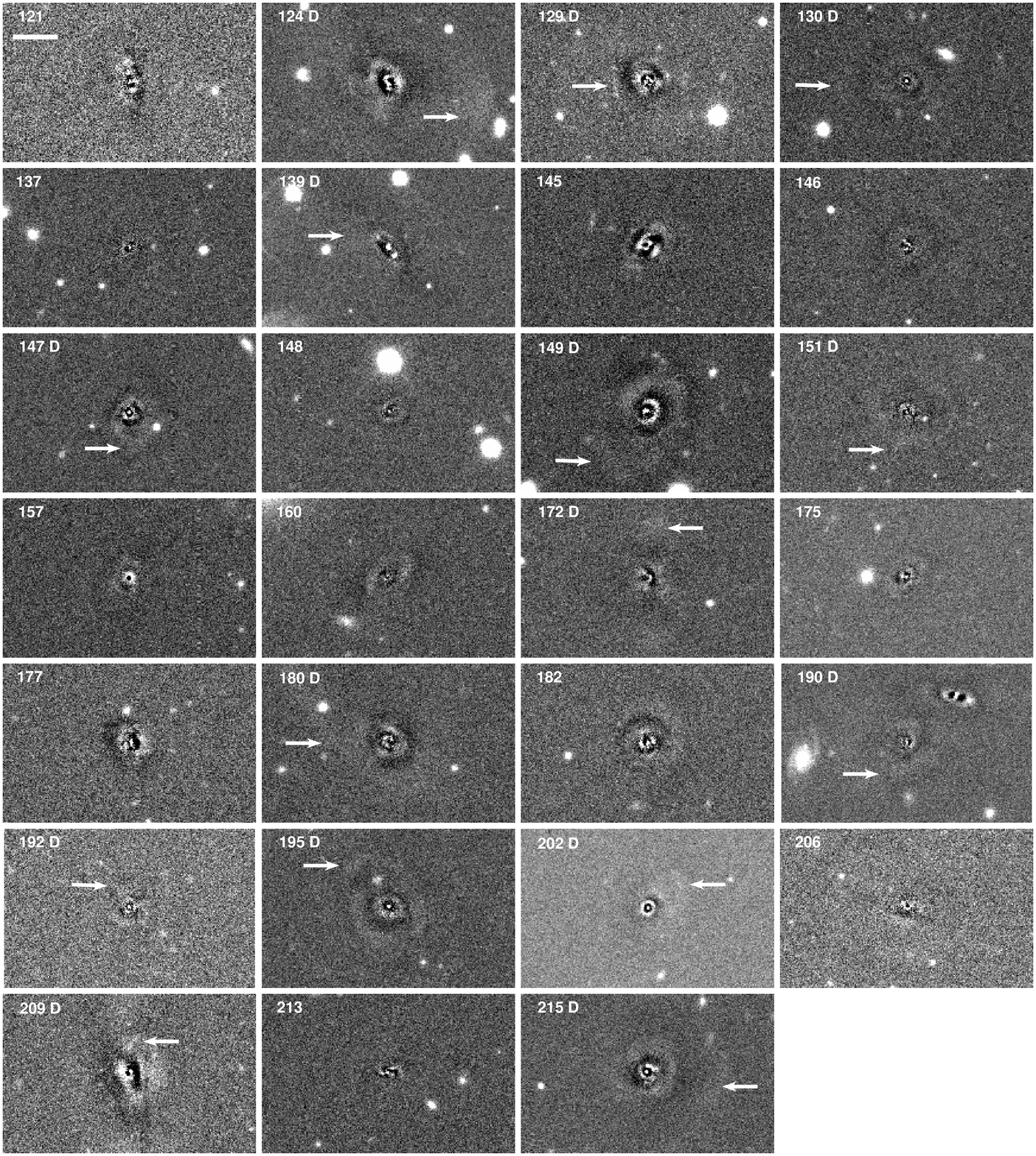}}\quad}
  \caption{Residuals of Sersic + exponential fit to SDSS $r$-band images of 55 star-forming blue ETGs from \citet{meert_2015}. The S09 object identification number is given at the top of each image. The
 galaxy residual images showing the presence of recent interaction features are marked with D, indicating the presence of stellar debris. All images are of the same dimension, and a line of length 15$\arcsec$ is shown in the first image; this corresponds to the typical size of the galaxy. The location of the stellar debris feature is indicated with an arrow.}
  \label{fig:fig2}
\end{figure*}

The SDSS $r$-band imaging data of 55 star-forming blue ETGs in the blue cloud were visually inspected for any signs of recent interaction. We found that many galaxies show features at high galactic radii
 that are indicative of recent interactions. Figure~\ref{fig:fig1} shows the $r$-band images of
 55 blue ETGs. The galaxies with objID 41, 50,
56, 61, 72, 124, 129, 149, 172, 195, and 215 show features at the galaxy outskirts in $r$-band images taken with 54s integration time. It is our interest to understand 
the origin of these features, which can shed light on the formation
 of the blue ETGs. The visual analysis of $r$-band images was further corroborated with the residuals of surface brightness profile fits on the galaxy images. We 
analysed the difference image obtained from the best-fit model and the galaxy image, which is the residuals to the profile fits. The
 visual analysis of the residual images of 55 star-forming galaxies clearly shows the deviation from a smooth featureless galaxy surface brightness profile typical
 of elliptical galaxies.  Figure~\ref{fig:fig2} shows the residuals from a Sersic $+$ exponential profile fit to the SDSS
 $r$-band images as explained in \citet{meert_2015}.  Analysing the residual image, we found that in addition to signatures of recent interactions, there are indications of deviations from 
a smooth featureless early-type morphology with galaxies hosting tidal tails, shells, asymmetric excess light far of and near
to the galactic centre, which we collectively call stellar debris.  We note that
 the residual images show a point source at the central region of the galaxy 
along with a ring structure, which demonstrates the failure of a simple two-component model to fit the underlying nature of the surface brightness profiles of blue ETGs. The normal ETGs 
from the catalogue of \citet{meert_2015} do not show such features in the residual images, as demonstrated in Fig. 2 of \citet{meert_2013}. Blue ETGs appear to have more structural
 inhomogenities than normal ETGs. We also checked the residual images from a single Sersic component
 fit to the galaxy surface brightness distribution, and the interaction features remains, regardless of the number of components used to model the galaxy surface brightness profile. 
We found that out of 55 blue ETGs, 32 galaxies show features that are indicative of recent merger events in the residual images, which we interpret as stellar debris. 
Table~\ref{galaxy details} provides the details of the structural  analysis based on visual inspection of the residual images along with details of the galaxy $u-r$ colours and star formation rates.  
We note that deep imaging of these galaxies is expected to reveal low surface brightness features that are otherwise not recorded in shallow SDSS images taken with 54s integration time.\\

\citet{Nair_2010} presented a detailed study of the morphology and the presence of fine structures (including interaction signatures such as tails, shells, and collisional rings) based
 on visual classifications for 14,034 galaxies in the Sloan Digital Sky Survey (SDSS)
Data Release 4 (DR4) for the redshift range, 0.01 < z < 0.1 and $g$ -band magnitude $<$ 16 mag. We note that 25 galaxies (brighter than $g$ $\sim$ 16 mag) in our sample of 55 blue ETGs
 are reported in their catalogue with galaxy details. Four galaxies are classified as with S0 and 21 with E morphology. The galaxy with ObjID 58 is classified with an Sa morphology. 
The number of galaxies listed with interaction features in their study are different (10) from our residual analysis (21), which
is  based on image decomposition. We note here in particular the galaxies with objIds 50 and
 215, which show signs of interaction in our residual analysis (Fig. 2), but are not reported in the analysis of \citet{Nair_2010}.
 
Galaxy Zoo classifications of blue ETGs are done purely based on visual morphology, and only those galaxies that secure $>$ 80 $\%$ votes are selected for
 the clean sample, as described in S09. This does not exclude S0 galaxies from our sample of 55 star-forming blue ETGs, and
 as described in S09, a few of our galaxies can have an S0 morphology. We call our sample early-type galaxies, which includes both ellipticals and S0s. The bias
 in the GZ classification process had been statistically studied by \citet{Bamford_2009}, who showed that contamination from other morphological types (particularly spirals)
  is 4$\%$, and for our star-forming blue ETGs, this translates into 2 galaxies.  It therefore follows that one of our galaxies is classified with an Sa morphology
 in the list of 25 blue ETGs from \citet{Nair_2010}.

The very recent galaxy morphology classification from Galaxy Zoo 2 supercedes the  original GZ
and compiles the morphologies for galaxies in the SDSS DR7 along with deeper images from Stripe 82 \citep{Willett_2013}. Fifty-two galaxies (from our sample of 55 star-forming blue ETGS) in the GZ2 catalogue 
 have a smooth morphology, which confirms the original GZ classification.
 
\subsection{Normal early-type galaxies}

The structural analysis of blue ETGs shows features that are indicative of recent interactions. The interesting question is how common these features are for 
normal ETGs on the red sequence of the galaxy colour-magnitude diagram. Tidal features are reported for ETGs on the red sequence for galaxies with an $r$-band absolute magnitude M$r$ $<$ -20.5 in the redshift range $z<0.05$ by visual inspection of the SDSS Stripe82 imaging data by \citet{Kaviraj_2010}. The SDSS Stripe 82 imaging is {$\sim$ 2mag} deeper than the normal SDSS r-band imaging and hence is an ideal data set in which to search for faint features that are indicative 
of recent interactions around ETGs that are otherwise not seen in the shallow SDSS imaging (and which can be revealed without an image decomposition technique).  The galaxies reported by 
\citet{Kaviraj_2010} consist of  238 relaxed ETGs without any morphological disturbances and 67 ETGs with tidal features exhibiting
 shells, fans, tidal tails, and dust lanes at the depth of the Stripe82 images. The analysis based on the position of galaxies on the 
 BPT diagram \citep{Baldwin_1981} shows that 1.3 $\%$ of  the
relaxed ETGs and 4.5 $\%$ of the ETGs with tidal features 
and dust lanes host ongoing star formation. However, the ETG
populations with relaxed and the tidal features have 
a similar distributions in ($u - r$)  colour (median $\sim$ 2.5 mag), indicating that any triggered star formation that is due to the interaction is weak and did not change the optical colours.
The recent interaction may not have brought an amount of gas sufficient for star formation to change the colours of these galaxies, as in the case of star-forming blue ETGs. 
It is worth noting that only 28 $\pm$ 3 $\%$ of the galaxies from the deeper Stripe 82 imaging data of \citet{Kaviraj_2010} show features that are indicative
 of recent interaction, whereas 58 $\pm$ 7 $\%$ of blue ETGs (with normal SDSS imaging) have interaction features in residual analysis. The uncertainties are calculated from 
considering the galaxy sample following a binomial probability distribution. The tidal features are more common around massive galaxies (stellar mass $> 10^{10.5}$ M$\odot$), and
 galaxies on the red sequence are twice as likely to show tidal features than galaxies in the blue cloud \citep{Atkinson_2013}.
The presence of tidal features around galaxy images depends on the depth of the data, and as summarised in Table 3 of  \citet{Kim_2012}, deeper 
imaging data retrieve more tidal features around ETGs. We expect a similar trend for blue ETGs in data from deep imaging ($\sim$ 2mag lower than SDSS)
 surveys like the DECam Legacy Survey of the SDSS Equatorial Sky (DECaLS). It has been demonstrated that the probability of finding features around blue ETGs is in general higher than for normal ETGs with
 deeper imaging data.

\begin{sidewaystable*}
\centering
\tabcolsep=0.05cm
\caption{\label{t7}Details of the structural analysis of 55 Blue early-type galaxies}\label{galaxy details} 
\fontsize{7}{8}\selectfont 
\setlength\LTleft{0pt}            
\setlength\LTright{0pt}           
\begin{longtable}{ccccccccccccccccccccccc} 
\hline\hline
S09& SDSS & RA & DEC  & $z$  &   M$_r$ &  $u-r$ &  SFR &  $<\mu_{e}>$ & $<e\mu_{e}>$   &  $<\mu_{e}>$ & $<e\mu_{e}>$ & $R_{e}$ &  e$R_{e}$ & $R_{e}$ &  e$R_{e}$ & $n$ & $en$  & $n$ & $en$ & $b/a$ & $b/a$ & Residual  \\ 
   &    &        &       &   &        &        &                 & ${ser}$        &    &  ${ser+exp}$      &  & ${ser}$ &   & ${ser+exp}$ &  & ${ser}$ &  & ${ser+exp}$ & & ${ser}$ & ${ser+exp}$ &  \\ 
 ID & ID & h:m:s  & d:m:s &   &  (mag) &  (mag) &  (M$\odot$/yr)  & (mag/arcsec$^2$)&    & (mag/arcsec$^2$) &  & (kpc)   &   & (kpc)       &  &         &  &             & &         &             & feature\\
 \hline 
  2    & 587722982272925748 & 11:23:27.0 & -00:42:48.8 & 0.04084 & -20.81 & 1.606 & 4.5  & 19.46 & 0.04 & 17.26 & 0.15      &     2.26 & 0.02 & 0.38  & 0.01    & 3.15 & 0.03 & 0.73 & 0.15    & 0.97 & 0.75 & Stellar debris \\
  4    & 587722982299271230 & 15:23:47.1 & -00:38:23.0 & 0.03747 & -21.0  & 1.998 & 2.5  & 19.48 & 0.05 & 17.79 & 0.02      &     2.32 & 0.03 & 0.65  & 0.0     & 4.22 & 0.04 & 1.03 & 0.01    & 0.85 & 0.83 &   \\
  5    & 587722983351582785 & 12:08:23.5 & +00:06:37.0 & 0.04081 & -21.5  & 2.038 & 3.5  & 19.57 & 0.04 & 18.11 & 0.05      &     3.37 & 0.03 & 1.1   & 0.01    & 3.67 & 0.02 & 1.56 & 0.02    & 0.91 & 0.93 &   \\
  7    & 587724198282133547 & 01:41:43.2 & +13:40:32.8 & 0.04539 & -21.81 & 1.432 & 12.0 & 18.58 & 0.01 & 18.6  & 0.01      &     2.15 & 0.01 & 1.64  & 0.0     & 0.97 & 0.01 & 0.4  & 0.0     & 0.86 & 0.81 &  Stellar debris \\
  8    & 587724199351812182 & 01:03:58.7 & +15:14:50.1 & 0.04176 & -21.4  & 2.261 & 7.0  & 20.58 & 0.08 & 20.14 & 0.24      &     5.1  & 0.08 & 3.37  & 0.16    & 4.87 & 0.04 & 5.36 & 0.12    & 0.71 & 0.7   & Stellar debris  \\
  14   & 587725550139408424 & 12:35:02.6 & +66:22:33.4 & 0.04684 & -21.67 & 1.959 & 6.1  & 19.23 & 0.11 & 16.62 & 0.04      &     3.62 & 0.08 & 0.71  & 0.01    & 8.0  & 0.11 & 2.3  & 0.04    & 0.81 & 0.75 &  Stellar debris \\
  20   & 587725981226107027 & 08:29:09.1 & +52:49:06.9 & 0.04842 & -21.0  & 1.835 & 2.1  & 19.9  & 0.09 & 19.95 & 0.0       &     3.06 & 0.05 & 3.13  & 0.0     & 3.55 & 0.05 & 3.64 & 0.0     & 0.95 & 0.95 &   \\
  23   & 587726031175548968 & 12:06:47.2 & +01:17:09.8 & 0.04124 & -21.18 & 2.207 & 1.2  & 19.61 & 0.06 & 20.36 & 0.29      &     3.0  & 0.04 & 4.13  & 0.24    & 4.18 & 0.04 & 6.53 & 0.28    & 0.74 & 0.73 &  \\
  30   & 587726100953432183 & 15:17:19.7 & +03:19:18.9 & 0.03749 & -20.74 & 2.131 & 0.73 & 20.41 & 0.02 & --    & --        &     3.2  & 0.02 & --   & --       & 1.09 & 0.01 & --   & --      & 0.71 & -- &   \\
  41   & 587728879266562060 & 10:16:28.4 & +03:35:02.7 & 0.04848 & -21.72 & 2.38  & 6.1  & 20.65 & 0.07 & 18.63 & 0.2       &     6.48 & 0.09 & 1.46  & 0.06    & 4.68 & 0.04 & 2.75 & 0.09    & 0.7  & 0.64 & Stellar debris  \\
  50   & 587729160046510114 & 13:57:07.5 & +05:15:06.8 & 0.03967 & -21.86 & 2.158 & 6.6  & 20.59 & 0.08 & 18.39 & 0.14      &     7.27 & 0.11 & 1.8   & 0.05    & 7.27 & 0.05 & 3.93 & 0.07    & 0.95 & 0.94 &  Stellar debris \\ 
  52   & 587729227683659793 & 14:51:15.7 & +62:00:14.6 & 0.04306 & -21.45 & 2.173 & 3.9  & 19.0  & 0.02 & 23.38 & 1.38      &     2.39 & 0.01 & 15.83 & 4.37    & 1.6  & 0.01 & 8.0  & 0.84    & 0.76 & 0.86 &  \\
  56   & 587729408622723373 & 17:23:24.9 & +27:48:46.3 & 0.04845 & -21.46 & 2.017 & 3.2  & 19.13 & 0.03 & 20.65 & 0.32      &     2.53 & 0.02 & 4.9   & 0.31    & 2.69 & 0.02 & 5.36 & 0.23    & 0.89 & 0.86 &  Stellar debris \\
  58   & 587729778516820024 & 14:06:56.4 & -01:35:41.0 & 0.02916 & -21.37 & 2.127 & 5.7  & 19.08 & 0.01 & 20.55 & 0.1       &     2.28 & 0.01 & 3.54  & 0.07    & 1.59 & 0.0  & 2.63 & 0.05    & 0.51 & 0.61 &   \\
  61   & 587730816286785593 & 22:15:16.2 & -09:15:47.6 & 0.03843 & -21.61 & 1.707 & 21.0 & 20.97 & 0.1  & 20.28 & 0.24      &     7.59 & 0.15 & 3.96  & 0.19    & 6.25 & 0.06 & 8.0  & 0.18    & 0.86 & 0.78 &  Stellar debris \\
  66   & 587731500261834946 & 10:54:37.9 & +55:39:46.0 & 0.04787 & -20.89 & 1.976 & 6.5  & 20.16 & 0.05 & 19.63 & 0.17      &     3.27 & 0.03 & 2.01  & 0.07    & 2.31 & 0.02 & 1.66 & 0.04    & 0.69 & 0.66 &  Stellar debris \\
  68   & 587731512617402577 & 03:01:26.2 & -00:04:25.5 & 0.04285 & -21.1  & 2.156 & 3.2  & 19.76 & 0.05 & 18.95 & 0.08      &     2.67 & 0.03 & 1.39  & 0.02    & 2.81 & 0.03 & 1.58 & 0.03    & 0.58 & 0.56  & Stellar debris \\
  72   & 587731521207730241 & 09:13:23.7 & +43:58:34.2 & 0.04292 & -21.45 & 2.176 & 14.0 & 20.41 & 0.07 & 20.68 & 0.2       &     5.18 & 0.07 & 5.51  & 0.21    & 4.56 & 0.04 & 5.71 & 0.14    & 0.72 & 0.68 &  Stellar debris  \\
  77   & 587731889505632389 & 12:20:37.4 & +56:28:46.2 & 0.04381 & -20.84 & 2.238 & 0.53 & 19.55 & 0.04 & 20.11 & 0.2       &     2.37 & 0.02 & 2.7   & 0.11    & 2.34 & 0.02 & 3.5  & 0.19    & 0.98 & 0.96 &   \\
  84   & 587732157389013064 & 07:54:20.6 & +25:51:33.2 & 0.04167 & -21.06 & 2.011 & 1.7  & 20.43 & 0.14 & 17.08 & 0.14      &     4.56 & 0.13 & 0.54  & 0.01    & 8.0  & 0.12 & 2.71 & 0.11    & 0.86 & 0.81 &  Stellar debris \\
  86   & 587732471463411876 & 08:53:11.4 & +37:08:06.5 & 0.0498  & -21.59 & 2.063 & 10.0 & 19.39 & 0.03 & 18.31 & 0.06      &     3.03 & 0.02 & 1.05  & 0.01    & 2.58 & 0.01 & 1.11 & 0.02    & 0.95 & 0.9  & Stellar debris \\
  92   & 587732769982906490 & 12:20:23.1 & +08:51:37.1 & 0.04894 & -20.88 & 2.086 & 3.6  & 19.63 & 0.04 & 19.19 & 0.06      &     2.45 & 0.02 & 1.65  & 0.02    & 2.05 & 0.02 & 1.34 & 0.02    & 0.67 & 0.62  & Stellar debris \\
  101  & 587733411518808182 & 14:02:48.8 & +52:30:00.8 & 0.04361 & -20.77 & 1.893 & 1.4  & 20.31 & 0.03 & 21.35 & 0.32      &     3.19 & 0.02 & 2.86  & 0.17    & 1.17 & 0.01 & 2.03 & 0.16    & 0.65 & 0.57  &   \\
  102  & 587733412055941186 & 14:07:47.2 & +52:38:09.7 & 0.04381 & -20.78 & 1.812 & 4.2  & 20.13 & 0.05 & 19.49 & 0.11      &     3.11 & 0.03 & 1.86  & 0.04    & 2.76 & 0.02 & 1.98 & 0.03    & 0.69 & 0.62   & Stellar debris \\
  103  & 587733412064788620 & 15:50:00.5 & +41:58:11.2 & 0.03391 & -20.8  & 1.879 & 3.6  & 19.76 & 0.02 & 23.23 & 1.13      &     2.56 & 0.01 & 11.87 & 2.69    & 2.01 & 0.01 & 8.0  & 0.75    & 0.89 & 0.88    &   \\
  105  & 587733432459788474 & 16:51:16.7 & +28:06:52.5 & 0.04724 & -21.59 & 1.921 & 4.0  & 19.68 & 0.03 & 22.13 & 0.71      &     3.42 & 0.02 & 9.18  & 1.29    & 2.43 & 0.02 & 8.0  & 0.55    & 0.55 & 0.7    &  Stellar debris \\
  108  & 587734622698602943 & 07:47:23.1 & +22:20:41.3 & 0.04549 & -20.9  & 2.256 & 0.45 & 21.03 & 0.08 & 19.0  & 0.1       &     4.89 & 0.08 & 0.68  & 0.01    & 2.81 & 0.03 & 0.83 & 0.06    & 0.85 & 0.82  &   \\
  119  & 587735742615388296 & 15:53:35.6 & +32:18:20.6 & 0.04985 & -21.07 & 1.789 & 3.2  & 20.43 & 0.13 & 19.49 & 0.33      &     4.52 & 0.12 & 2.24  & 0.15    & 6.25 & 0.09 & 6.81 & 0.25    & 0.73 & 0.63   &  Stellar debris \\
  121  & 587736477586554950 & 13:47:47.7 & +11:16:27.0 & 0.03942 & -21.2  & 1.911 & 5.5  & 19.78 & 0.05 & 19.58 & 0.07      &     3.21 & 0.03 & 2.68  & 0.04    & 3.72 & 0.03 & 3.75 & 0.04    & 0.6  & 0.55    &   \\
  124  & 587736584980463705 & 16:44:30.8 & +19:56:26.7 & 0.023   & -20.71 & 1.873 & 5.2  & 19.96 & 0.04 & 19.5  & 0.09      &     2.8  & 0.02 & 1.81  & 0.03    & 4.25 & 0.02 & 4.75 & 0.05    & 0.85 & 0.72    &  Stellar debris \\
  129  & 587738067267878973 & 07:59:12.4 & +53:33:26.0 & 0.03479 & -20.92 & 1.657 & 13.0 & 18.98 & 0.04 & 17.33 & 0.02      &     1.93 & 0.02 & 0.55  & 0.0     & 4.19 & 0.03 & 0.89 & 0.01    & 0.79 & 0.8     &  Stellar debris  \\ 
  130  & 587738067269255432 & 08:10:20.1 & +56:12:26.3 & 0.04623 & -20.78 & 2.13  & 1.0  & 20.25 & 0.06 & 19.3  & 0.1       &     3.1  & 0.04 & 1.42  & 0.03    & 2.81 & 0.03 & 1.39 & 0.03    & 0.9  & 0.93    &  Stellar debris \\
  137  & 587738947194519672 & 07:56:36.3 & +18:44:17.7 & 0.03988 & -21.53 & 2.088 & 3.9  & 20.4  & 0.06 & 19.62 & 0.23      &     5.25 & 0.06 & 2.89  & 0.13    & 4.79 & 0.03 & 4.4  & 0.11    & 0.72 & 0.71     &   \\
  139  & 587739115234394179 & 07:56:08.7 & +17:22:50.5 & 0.02899 & -20.73 & 2.211 & 1.6  & 19.99 & 0.03 & 22.48 & 0.5       &     2.83 & 0.02 & 7.45  & 0.74    & 2.6  & 0.01 & 8.0  & 0.35    & 0.46 & 0.6     &  Stellar debris \\
  145  & 587739406262468657 & 15:18:09.6 & +25:42:11.5 & 0.0326  & -20.85 & 1.96  & 6.6  & 20.46 & 0.06 & 17.84 & 0.07      &     3.81 & 0.04 & 0.49  & 0.01    & 4.05 & 0.03 & 1.73 & 0.05    & 0.81 & 0.66    &   \\
  146  & 587739406268039264 & 16:07:54.0 & +20:03:03.8 & 0.03165 & -20.73 & 1.487 & 4.8  & 20.56 & 0.15 & 17.28 & 0.03      &     4.02 & 0.12 & 0.48  & 0.0     & 8.0  & 0.12 & 1.17 & 0.02    & 0.95 & 0.84     &   \\
  147  & 587739506086707246 & 13:26:20.8 & +31:41:59.9 & 0.04999 & -21.89 & 1.93  & 6.7  & 19.76 & 0.02 & 18.54 & 0.06      &     4.27 & 0.02 & 1.02  & 0.01    & 2.21 & 0.01 & 0.96 & 0.03    & 0.84 & 0.78     &  Stellar debris \\
  148  & 587739647814139970 & 10:20:34.9 & +29:14:10.8 & 0.04846 & -20.96 & 1.998 & 1.0  & 19.68 & 0.03 & 24.24 & 2.67      &     2.55 & 0.01 & 15.85 & 8.45    & 1.3  & 0.01 & 6.02 & 1.33    & 0.82 & 0.85     &   \\
  149  & 587739648357826596 & 11:31:22.0 & +32:42:22.9 & 0.03368 & -21.61 & 2.015 & 6.3  & 20.11 & 0.04 & 18.73 & 0.16      &     4.89 & 0.04 & 1.77  & 0.06    & 4.76 & 0.02 & 3.6  & 0.07    & 0.79 & 0.79     &  Stellar debris \\
  151  & 587739810496512078 & 15:44:51.5 & +17:51:22.5 & 0.04521 & -21.32 & 1.994 & 3.7  & 19.82 & 0.03 & 20.35 & 0.13      &     3.36 & 0.02 & 4.11  & 0.1     & 2.67 & 0.01 & 3.48 & 0.08    & 0.83 & 0.86     &  Stellar debris \\
  157  & 587739828749730189 & 16:04:39.4 & +16:44:43.6 & 0.04599 & -20.74 & 2.308 & 0.54 & 20.74 & 0.04 & --    & --        &     3.85 & 0.03 & --    & --      & 1.77 & 0.01 & --   & --      & 0.61 & --     &   \\
  160  & 587741490904105107 & 10:25:24.7 & +27:25:06.3 & 0.04973 & -20.98 & 2.256 & 3.1  & 21.09 & 0.12 & 17.67 & 0.1       &     5.83 & 0.14 & 0.43  & 0.01    & 4.57 & 0.06 & 0.59 & 0.08    & 0.54 & 0.6      &   \\
  172  & 588007005234856197 & 08:17:56.3 & +47:07:19.5 & 0.03901 & -21.06 & 2.309 & 8.1  & 20.87 & 0.07 & 19.95 & 0.37      &     4.98 & 0.07 & 2.08  & 0.15    & 3.55 & 0.03 & 3.67 & 0.15    & 0.82 & 0.92     &  Stellar debris \\
  175  & 588010135730782222 & 09:24:29.6 & +53:41:37.8 & 0.0459  & -20.8  & 1.997 & 0.63 & 19.7  & 0.03 & 19.59 & 0.11      &     2.48 & 0.02 & 2.18  & 0.05    & 2.09 & 0.02 & 1.97 & 0.03    & 0.87 & 0.86     &   \\
  177  & 588010878765826140 & 13:01:41.4 & +04:40:49.9 & 0.03836 & -21.4  & 2.101 & 1.5  & 19.36 & 0.03 & 20.42 & 0.37      &     2.85 & 0.02 & 3.42  & 0.25    & 2.83 & 0.01 & 8.0  & 0.34    & 0.89 & 0.9      &   \\
  180  & 588011125186691149 & 12:06:17.0 & +63:38:19.0 & 0.03974 & -21.26 & 1.686 & 18.0 & 19.26 & 0.02 & 18.84 & 0.03      &     2.46 & 0.01 & 1.13  & 0.01    & 1.75 & 0.01 & 0.52 & 0.01    & 0.83 & 0.64     &  Stellar debris \\
  182  & 588011216991289349 & 14:32:22.7 & +56:51:08.4 & 0.04302 & -21.74 & 1.864 & 6.1  & 19.27 & 0.02 & 18.35 & 0.03      &     3.14 & 0.01 & 1.3   & 0.01    & 2.53 & 0.01 & 0.93 & 0.01    & 0.98 & 0.97     &   \\
  190  & 588017604156784724 & 14:14:33.2 & +40:45:22.9 & 0.04185 & -20.86 & 1.614 & 5.2  & 19.15 & 0.06 & 17.8  & 0.08      &     2.13 & 0.02 & 0.83  & 0.01    & 4.46 & 0.05 & 1.8  & 0.04    & 0.78 & 0.81     &  Stellar debris \\
  192  & 588017606292865048 & 11:52:05.0 & +45:57:06.6 & 0.04316 & -20.72 & 1.842 & 1.7  & 18.97 & 0.04 & 18.09 & 0.04      &     1.71 & 0.01 & 0.76  & 0.01    & 2.86 & 0.03 & 0.89 & 0.02    & 0.91 & 0.93     &  Stellar debris\\
  195  & 588017627778580560 & 14:53:23.4 & +39:04:13.6 & 0.03153 & -20.88 & 2.134 & 1.2  & 19.96 & 0.06 & 19.6  & 0.15      &     3.28 & 0.04 & 2.43  & 0.07    & 5.4  & 0.04 & 5.63 & 0.09    & 0.84 & 0.85     &  Stellar debris \\
  202  & 588017978903232631 & 14:17:32.6 & +36:20:19.1 & 0.04712 & -20.93 & 1.916 & 1.6  & 19.75 & 0.04 & 19.71 & 0.06      &     2.8  & 0.02 & 2.73  & 0.03    & 2.83 & 0.02 & 2.79 & 0.02    & 0.9  & 0.9      &  Stellar debris \\
  206  & 588017991233110227 & 14:37:33.0 & +08:04:43.0 & 0.04987 & -21.62 & 2.353 & 3.6  & 20.74 & 0.1  & 21.08 & 0.31      &     6.79 & 0.13 & 7.06  & 0.44    & 5.31 & 0.06 & 7.25 & 0.2     & 0.59 & 0.57     &   \\
  209  & 588018254297038899 & 16:18:18.7 & +34:06:40.1 & 0.04733 & -21.58 & 2.307 & 2.9  & 21.03 & 0.09 & 20.18 & 0.36      &     7.74 & 0.14 & 3.62  & 0.26    & 5.75 & 0.06 & 7.83 & 0.25    & 0.8  & 0.74     &  Stellar debris \\
  213  & 588297863102988421 & 08:43:46.7 & +31:34:52.6 & 0.04756 & -20.7  & 2.11  & 7.3  & 19.25 & 0.03 & 19.44 & 0.19      &     1.82 & 0.01 & 0.65  & 0.02    & 1.26 & 0.01 & 0.64 & 0.09    & 0.57 & 0.45     &   \\
  215  & 588848899365601360 & 10:26:54.6 & -00:32:29.4 & 0.03463 & -21.38 & 2.137 & 7.6  & 19.56 & 0.04 & 18.61 & 0.14      &     3.04 & 0.02 & 1.55  & 0.04    & 4.06 & 0.02 & 2.77 & 0.05    & 0.97 & 0.97       &  Stellar debris\\
  \hline
\end{longtable}
\tablefoot{Col. (1): object ID for the galaxies as in S09; Col. (2): SDSS DR7 ObjID; Cols. (3) and (4): galaxy coordinates (epoch J2000); Col. (5): spectroscopic redshift of blue ETGs from SDSS; Col. (6):  
$r$-band absolute magnitude; Col. (7) : $u-r$ colors; Col. (8): star formation rates ; Cols. (9) and (10):
computed galaxy mean surface brightness and the error in computation for Sersic only; Cols. (11) and (12):
 computed galaxy mean surface brightness and the error in computation for Sersic + exponential; Cols. (13) and (14): effective radii in Kpc and the error for Sersic only; Cols. (15) and (16): effective radii in Kpc and the error for Sersic
 + exponential; Cols. (17) and (18): Sersic index and error for Sersic only; Cols. (19) and (20): Sersic index and error for Sersic + exponential; Cols. (21): axial ratio for Sersic only; Cols. (22): axial ratio for Sersic + exponential; Col. (23): detail obtained
 from the visual inspection of the residuals of a GALFIT analysis of an SDSS galaxy $r$-band image.
}
\end{sidewaystable*}

\pagebreak

\subsection{Enhanced star formation rate and presence of tidal debris}

\begin{figure*}
\includegraphics[width=10.0cm,height=10.0cm,keepaspectratio]{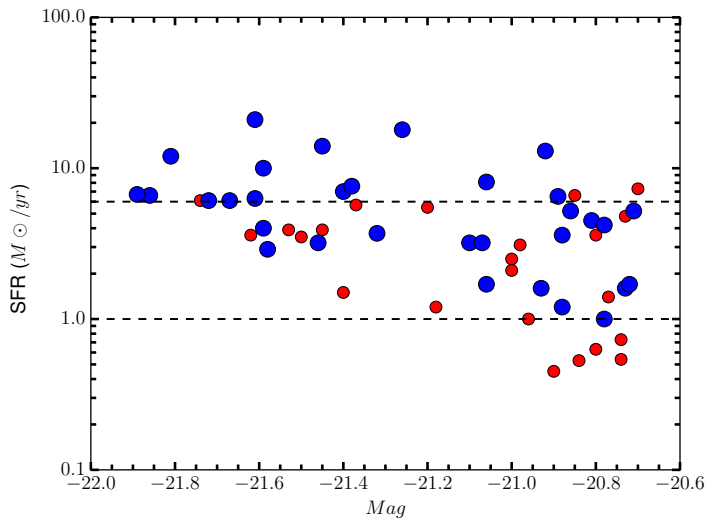}
\caption{Star formation rates of 55 star-forming blue ETGs plotted against the galaxy $r$-band absolute magnitude. Galaxy with interaction features are shown in blue
and those without detected features in red. Galaxies are separated into three SFR bins  using the dotted lines at 1 and 6  M$\odot$/yr.}
\label{fig:fig3}
\end{figure*}

The presence of stellar debris in the residual images of galaxies can be interpreted as an indication of recent galaxy-scale interactions. The detailed knowledge 
of the mass ratios, stellar population content, and the gas exchanged during these galaxy-scale interaction is crucial for our understanding of the formation and evolution
of blue ETGs. We are interested in knowing the influence of recent interaction on the ongoing star formation in blue ETGs. We therefore investigated the star formation rates of the blue ETGs with 
and without stellar debris in the residual images. We performed a Kolmogorov-Smirnov (KS) test on the star formation rates of galaxies with and with out stellar debris in residual images and found that
 with a 95.73 $\%$ probability, they are drawn from two different galaxy distributions. This clearly indicates that galaxies with stellar debris show a different trend in star formation rate from galaxies without detected features.
The $r$-band absolute magnitude (based on SDSS petromag, which measures the total light of the galaxy) of blue ETGs is plotted against
 the star formation rates in Fig.~\ref{fig:fig3}. Blue ETGs with stellar debris identified in the residual analysis are shown in blue and those without detected features in red. We divided the galaxies into three bins of star formation rates: SFR
 $<$ 1  , 1 to 6, and $>$ 6 M$\odot$/yr.  Blue ETGs with SFR $<$ 1 M$\odot$/yr show no features that would be indicative of recent interaction (5 galaxies), whereas galaxies of SFR 1 to 6 M$\odot$/yr are a 
 mix of galaxies with (17 galaxies) and without (15 galaxies) tidal features. Finally, the $>$ 6 M$\odot$/yr region is dominated (15/18) by galaxies with tidal interaction features.
 We note that there is a general trend for galaxies with higher luminosity to display interaction signatures and  high star formation rates, as confirmed with a KS test. The high star formation rates 
 are responsible for the higher luminosity of these galaxies, and based on the position of galaxies in the SFR plot, we
 argue that the recent interaction might have brought in the sufficient fuel (cold molecular hydrogen gas or molecular gas) for the observed intense star formation rates
  in an otherwise normal elliptical galaxy. The galaxies with high star formation rates might have acquired the gas very recently, where the intense ongoing 
star formation makes the galaxies both bluer and brighter. The SFR and the absolute magnitude will decrease and the colour will increase depending on the depletion of molecular gas (once acquired from interaction) and will decide the fate of the
 blue ETG.\\

\section{Scaling relations of blue early-type galaxies}

The Kormendy relation  is a scaling relation observed between the effective radii $R_{e}$ (kpc) and 
the mean surface brightness at the effective radius $<\mu_{e}>$  (mag/arcsec$^2$) of ETGs \citep{Kormendy_1977}. 
The ETGs in all environments are found to follow this tight relation up to redshift $\sim$ 1 \citep{Labarbera_2003,Reda_2004,Diserego_2005}.
The circularised effective radius is first computed from the measured effective radius ($a_{e}$) and axial ratio ($b/a$) using the equation $R_{e} = a_{e}\sqrt{b/a}$.
The mean surface brightness is calculated from the circularised effective radius
 ($arcsec$), total magnitude ($mag$), and redshift ($z$) using the following equation:\\

$<\mu_{e}>$ = $mag$ + 2.5log(2$\pi$$R_{e}^2$) - 10log(1+$z$). \\

The mean surface brightness ($<\mu_{e}>$), effective radii $R_{e}$ (kpc), axial ratio ($b/a$), and total magnitude ($mag$)
 is used to construct the scaling relations of blue ETGs.

\subsection{Double-component case: Sersic $+$ exponential profile}

\begin{figure*}
\centering
\includegraphics[width=15.0cm,height=15.0cm,keepaspectratio]{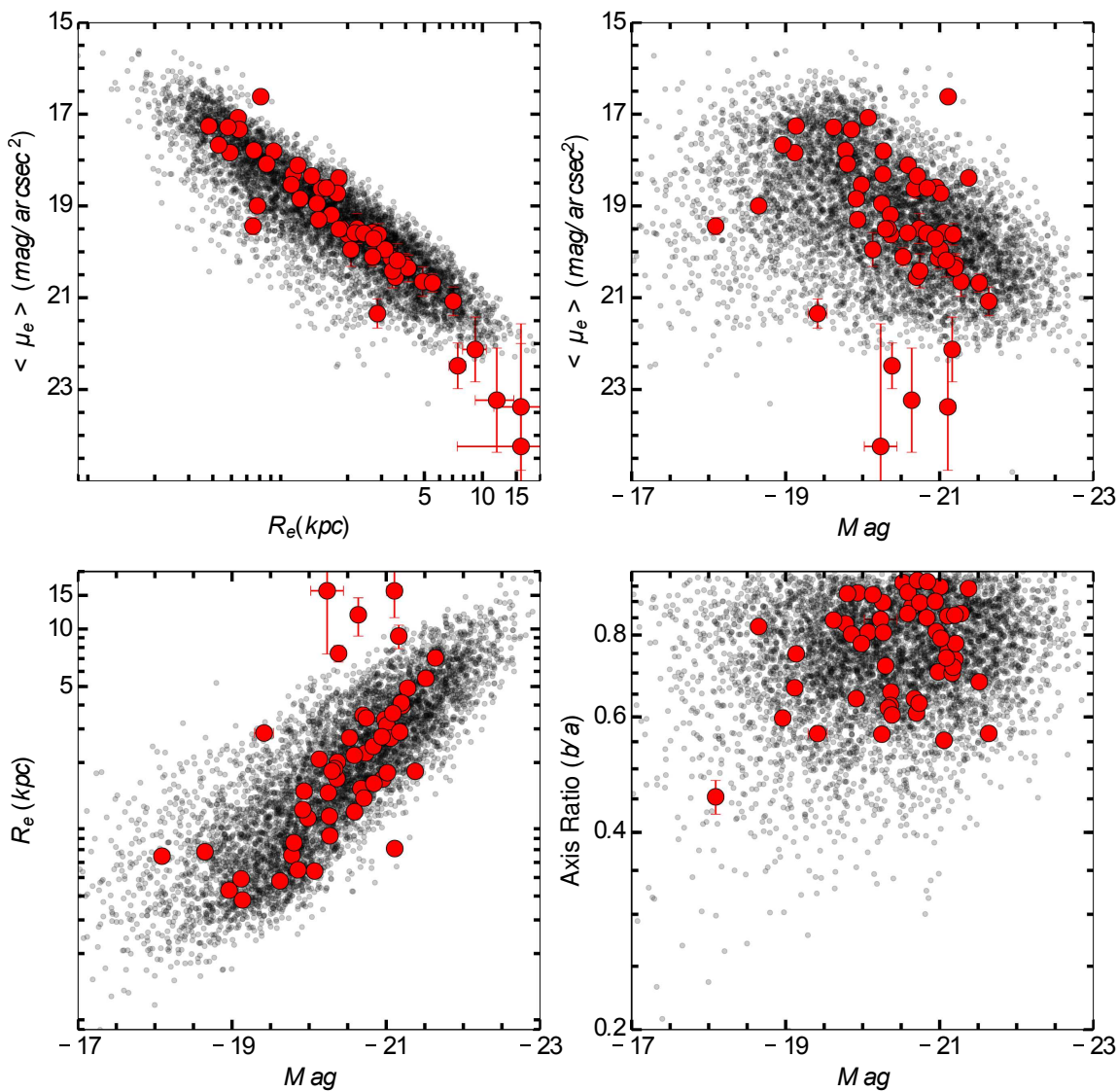}
\caption{55 star-forming blue ETGs on the galaxy scaling relations. The
 parameters used in the diagram are derived from the $r$-band SDSS imaging data. Grey points are normal elliptical galaxies in the redshift range 0.02 $>$ $z$ $>$ 0.05.}
\label{fig:fig4}
\end{figure*}

\begin{figure*}
\includegraphics[width=7.5cm,height=7.5cm,keepaspectratio]{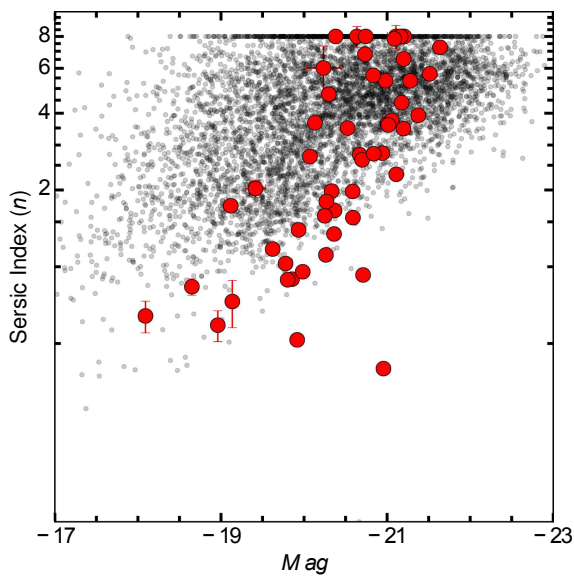}
\caption{Sersic index ($n$) of 55 star-forming blue ETGS plotted against the $r$-band absolute magnitude. The
 parameters used in the diagram are derived from the $r$-band SDSS imaging data. Grey points are normal elliptical galaxies in the redshift range 0.02 $>$ $z$ $>$ 0.05.}
\label{fig:fig5}
\end{figure*}

We constructed the Kormendy relation of star-forming blue ETGs using the measured structural parameters from a double-component model fit to the galaxy surface brightness distribution.
Figure~\ref{fig:fig4} shows the $r$-band structural parameters of 55 blue ETGs on the scaling relations. We also plot the scaling relation
 of normal elliptical galaxies in the redshift range 0.02 $>$ $z$ $>$ 0.05 from the Meert catalogue (in grey) along with the 55
 star-forming blue ETGs (red).  We selected only those 
galaxies that are morphologically ellipticals confirmed from GalaxyZoo \citep{Lintott_2008} to assemble the normal elliptical galaxy list. 
It is clear from figure that blue ETGs and normal elliptical galaxies share similar structural properties. The 
Sersic index of the 55 blue ETGs is shown in Fig.~\ref{fig:fig5}.\\
Two galaxies (objID 30,157) are having unrealistic errors in the measured structural parameters and hence we did not use them when we constructed the scaling relation.
We note that 5 galaxies (objId 52,103,105,139,148) have low surface brightness and high effective surface brightness for their measured absolute magnitude. These 
5 galaxies have high Sersic index values (saturating at $n$ $\sim$ 8).
There are 23 galaxies with Serisc index values $<$ 2, which implies
that they are disc systems. The size-magnitude relation of normal ellipticals and blue ellipticals shows the similar trend. Blue ETGs are
 bulge-dominated systems with $b/a$ values  greater than 0.5, which further confirms the morphological classification.  

\subsection{Single-component case: Sersic profile}

We also constructed the scaling relation using the galaxy structural parameters estimated with the single Sersic model fit to the
surface brightness distribution. The 55 blue ETGs fall on the scaling relation along 
 with normal elliptical galaxies.
We note that for the single Sersic component case, 9 galaxies have a Sersic
 index value $<$ 2 and 3 have a Sersic index saturating at index value 8. We checked 
the $r$-band images and the residual images of the low Sersic index galaxies and conclude with
 the following inferences. For the galaxies with objID 7 and 58, the galaxy image is in the corner of the fitting box region and the fitting process 
may be affected by this artefact. The galaxies with objID 30,
52, 157, 180, and 213 show features in the residual images that
are indicative of an 
intrinsic disc nature of the galaxy. The galaxy with objID 148 has a bright object nearby that might affect the fitting process.
The galaxy with objID 101 shows a patch in the residual images that might be due to the presence of dust lanes. The galaxies with a Sersic index value 8, with objId 14 
and 146, have a point source in the centre region of the residual images that is suggestive of a steep core, and the galaxy with objID 84 has bright objects nearby that affect the fitting process.\\

The main inferences from the scaling relation of star-forming blue ETGs are the following: blue ETGs obey the Kormendy relation and the size-magnitude relation of normal (red and dead) elliptical galaxies. The stellar 
population content  may be different (with blue colours and ongoing star formation), but the structure and the dynamical state of blue ETGs are comparable to normal elliptical galaxies in the local Universe. 
The double-component fits (Sersic $+$ exponential) shows galaxies with low Sersic index values following a shallow declining light distribution in contrast to the steep distribution  of the single-component case (see Table 1). 
We infer that this is due to the presence of a disc component in a good fraction (23/55) of blue ETGs where star formation might be occurring. The star formation
 in blue ETGs is expected to be due to recent events that occurred in a normal elliptical galaxy that did not alter the overall morphological structure of the galaxy. It is therefore worth
 investigating events in the local Universe that can cause star formation without disturbing the structure of a normal elliptical galaxy.\\

\section{Discussion}

The position of star-forming blue ETGs on the colour-magnitude/colour-stellar mass diagrams has been interpreted as possible migration from the red sequence to the blue cloud after acquiring sufficient
 fuel for star formation or fading post-starburst galaxies \citep{Kannappan_2009,Huertas_2010,Thilker_2010,Marino_2011,Moffett_2012,McIntosh14,Rutkowski_2014,Moffett_2015,Wong_2015}. The sample of a few blue ETGs
 from other studies was found to host a significant amount of molecular gas ($\sim 10^{7-9}$ M$\odot$) with a linear dependence on star formation
 surface density within the galaxy and a higher molecular gas star formation efficiency than in normal star-forming late-type galaxies \citep{Wei_2010,Stark_2013}. The
 higher efficiency in converting the molecular gas to stars is
expected to rapidly deplete the fuel for star formation, and in the absence of an external supply of fresh gas, the galaxy 
can rapidly change to a normal red and dead ETG. \citet{Stark_2013} analysed the relationship between mass-corrected blue centredness (tracing recent enhancements of central star formation) and
total molecular-to-atomic $(H_{2}/HI)$ gas ratios for early- to late-type field galaxies with stellar masses of $\sim 10^{7.2-11.2}$ M$\odot$. Blue ETGs are found to follow a different relation (compared to
 spiral galaxies, for which a positive correlation exists, which is interpreted as a result of local galaxy interactions and molecular gas supply) and are found to be late or starburst merger remnants with the starburst
 (keeping the centre blue) rapidly depleting the molecular gas. We consider the extreme case of our sample with star formation rate of 21 M$\odot$/yr holding molecular gas content of $\sim 10^{9}$ M$\odot$ as seen
 in other studies of blue ETGs \citep{Wei_2010,Stark_2013}. If we consider a 50$\%$ molecular gas star formation efficiency, then half of the molecular gas content will be converted into stars in 
$10^{8}$ yr. The gas content will be completely exhausted
 in $2 \times 10^{8}$ yr and will change the blue ETG back to a normal 
 galaxy on the red sequence (in $<$ 1 Gyr timescale). Blue ETGs would then be a transient phase in the evolution of a normal ETG and explain why they contribute only a
 low fraction (5.7 $\pm$ 0.4 $\%$) to the low-redshift ETG population.

We found  that 58 $\%$  of  the star-forming blue ETGs host signatures of recent interactions and fall, along with 
normal ellipticals, on the galaxy scaling relations.  This demonstrated that the sample of 55 galaxies has characteristics of normal elliptical galaxies except for the
high star formation rates and blue colours. We suggest that the star formation seen in these otherwise passively evolving red
and dead stellar systems is related to the recent interactions. The intense star formation found in these galaxies needs a 
huge reservoir of molecular gas (particularly for those galaxies with high star formation rates that reach as high as 21 M$\odot$/yr),  which is 
generally not available in intrinsically gas-poor elliptical galaxies \citep{Young_2011,Young_2014}. We invoke the scenario of a recent interaction with a gas-rich galaxy that could provide the
 fuel for intense star formation along with the imprint of interaction in the form of tidal tails and stellar debris. The nature
 of the interaction could be a major merger (1/1 in mass ratio), minor merger ($<$ 1/3 in mass ratio), or a fly-by, depending on the 
strength of the tidal features and the dynamical state of the galaxy.

The ETGs in the local Universe are known to host 
a low level of recent star formation {\citep{Trager_2000, Yi_2005}, and there is 
accumulating evidence that indicates that minor mergers are responsible for this low-level
 star formation {\citep{Kaviraj_2009, Kaviraj_2014}. The majority of ETGs
 in the local Universe are found to have signatures of recent interaction and a low level of recent
 star formation {\citep{Kaviraj_2010}. The recent interaction may
 have provided the fuel for the observed low level of star formation.
The reported ETGs in low-density environment with  
reservoirs of neutral hydrogen in regular rotating discs and extended morphologies  further
 support this argument \citep{Serra_2010}. Numerical simulations have also demonstrated the occurrence of gas rings and the accretion of fresh gas in
 ETGs from minor mergers \citep{Mapelli_2015, Kaviraj_2009, Peirani_2010}.

The formation of shells and tidal tails has been reproduced in N-body simulations of
 galaxy minor mergers \citep{Quinn_1984, Feldmann_2008}. \citet{Feldmann_2008} were
 able to reproduce the types of tidal features we observed here
with mass ratios of 1:10 between the accreting elliptical and the accreted disc galaxy. The authors argued that mergers
between a disc galaxy and an elliptical galaxy can form tidal features that last for $\sim$ 2 Gyr in comparison to mergers between equal-mass galaxies, which 
lasts for a few million years.

The connection between the star-forming blue ETG and polar-ring galaxies 
in the local Universe is worth investigating. A polar-ring galaxy system consists of a 
 gas-poor S0 galaxy with a gas-rich  star-forming polar ring held perpendicular
 to the semi-major axis of host galaxy \citep{Whitmore_1990}. The 
favoured scenario of the polar-ring formation is the accretion of a gas-rich dwarf galaxy by an ETG 
\citep{Bournaud_2003}. The polar ring of the  galaxy is found to be stable, young, and gas rich (molecular gas mass of the range
 $\sim 10^{10}$ M$\odot$) compared to the old, gas-poor host galaxy \citep{Galletta_1997,vanDriel_2000,Iodice_2002a,Iodice_2002b,Iodice_2002c,vanDriel_2002,Combes_2013}. 
We consider the case where the central host galaxy is replaced by a normal gas poor elliptical galaxy, which is then allowed to interact 
with a gas-rich dwarf galaxy. The system may not be stable like in the case of a polar-ring galaxy, but instead results in the formation of a
 gas-rich elliptical galaxy with intense star formation from the freshly
 acquired gas, showing signs of recent interaction in the form of tidal tails and shells. The end
  product of this scenario is a blue ETG with the characteristics of
the galaxies we studied here.

We have considered possible
triggers of star formation in an otherwise red and dead elliptical
galaxy. While we cannot point to one compelling source, we showed
signatures of recent interaction in images of galaxies with high star formation rates, 
 and we suggest that the interaction with a gas-rich companion
 could have triggered the star formation in elliptical galaxies. (We note that the interaction features have been detected around images of blue ETGs with low star formation
 rates and also around normal galaxies.) The interaction with the gas-rich companion may have provided the fuel for star formation. Star
 formation occurs in discs where gas settles and stabilises. We showed signatures of disc component in 23 blue ETGs from the structural analysis. The star-forming phase might be  transient ($<$ 1 Gyr), which
 is supported by the low fraction of such galaxies in the local Universe.  The intense star formation rate, in the absence of a continuous supply
 of molecular gas, will rapidly deplete the fuel for further star formation. Blue ETGs
 will then move from the blue cloud to the red sequence on the galaxy colour-magnitude diagram. The transient path in the evolution of a normal
 elliptical galaxy will then contribute to the build-up of the
stellar mass of
 ETGs in the local Universe.\\

\section{Summary}
  We presented a structural analysis of the SDSS $r$-band imaging data of  55 star-forming blue ETGs. We found that the
 residuals of the 32 galaxy surface brightness profile fits show structural features that are indicative of recent interactions. Galaxies with higher luminosity display 
interaction signatures and high star
formation rates. The star-forming blue ETGs follow the Kormendy relation and show characteristics of normal elliptical galaxies on the galaxy
 scaling relations. The star-forming population of blue ETGs at low redshifts could be 
normal ellipticals that may have undergone a recent minor-merger event. The star formation
 in these
 galaxies will shut down once the recently acquired fuel is consumed, and the galaxies will
 evolve into normal ETGs. 
 Blue ETGs are probably transient systems that are depleted of the fuel for star formation and become a 
normal ETG there by increasing the stellar mass on the red sequence in the local Universe.\\

\begin{acknowledgements}
We are grateful to the referee for critical comments that improved the quality of the paper. We thank Alan Meert for the blue early-type galaxy residual images and Sugata Kaviraj for the catalogue on peculiar early-type galaxies in the Sloan Digital Sky Survey Stripe82. Funding for the SDSS and SDSS-II has been provided by the Alfred P. Sloan Foundation, the Participating Institutions, the National Science Foundation, the U.S. Department of Energy, the National Aeronautics and Space Administration, the Japanese Monbukagakusho, the Max Planck Society, and the Higher Education Funding Council for England. The SDSS Web Site is http://www.sdss.org/.\\
The SDSS is managed by the Astrophysical Research Consortium for the Participating Institutions. The Participating Institutions are the American Museum of Natural History, Astrophysical Institute Potsdam, University of Basel, University of Cambridge, Case Western Reserve University, University of Chicago, Drexel University, Fermilab, the Institute for Advanced Study, the Japan Participation Group, Johns Hopkins University, the Joint Institute for Nuclear Astrophysics, the Kavli Institute for Particle Astrophysics and Cosmology, the Korean Scientist Group, the Chinese Academy of Sciences (LAMOST), Los Alamos National Laboratory, the Max-Planck-Institute for Astronomy (MPIA), the Max-Planck-Institute for Astrophysics (MPA), New Mexico State University, Ohio State University, University of Pittsburgh, University of Portsmouth, Princeton University, the United States Naval Observatory, and the University of Washington.
\end{acknowledgements}



\begin{thebibliography}{99}
\bibitem[Abazajian et al.(2009)]{Abazajian_2009} Abazajian, K.~N., 
Adelman-McCarthy, J.~K., Ag{\"u}eros, M.~A., et al.\ 2009, \apjs, 182, 543 

\bibitem[Atkinson et al.(2013)]{Atkinson_2013} Atkinson, A.~M., Abraham, R.~G., \& Ferguson, A.~M.~N.\ 2013, \apj, 765, 28 

\bibitem[{{Baldry} {et~al}\mbox{.}(2004){Baldry}, {Glazebrook}, {Brinkmann},
  {Ivezi{\'c}}, {Lupton}, {Nichol}, \& {Szalay}}]{Baldry_2004}
{Baldry} I.~K., {Glazebrook} K., {Brinkmann} J., {Ivezi{\'c}} {\v Z}., {Lupton}
  R.~H., {Nichol} R.~C., {Szalay} A.~S., 2004, \apj, 600, 681
\bibitem[Baldwin et al.(1981)]{Baldwin_1981} Baldwin, J.~A., 
Phillips, M.~M., \& Terlevich, R.\ 1981, \pasp, 93, 5 
\bibitem[Bamford et al.(2009)]{Bamford_2009} Bamford, S.~P., Nichol, R.~C., Baldry, I.~K., et al.\ 2009, \mnras, 393, 1324 

\bibitem[{{Bell} {et~al}\mbox{.}(2004){Bell}, {Wolf}, {Meisenheimer}, {Rix},
  {Borch}, {Dye}, {Kleinheinrich}, {Wisotzki}, \& {McIntosh}}]{Bell_2004}
{Bell} E.~F. {et~al.}, 2004, \apj, 608, 752
\bibitem[Bournaud \& Combes(2003)]{Bournaud_2003} Bournaud, F., \& Combes, F.\ 2003, \aap, 401, 817 
\bibitem[Brown et al.(2007)]{Brown_2007} Brown, M.~J.~I., Dey, A., 
Jannuzi, B.~T., et al.\ 2007, \apj, 654, 858 
\bibitem[Combes et al.(2013)]{Combes_2013} Combes, F., Moiseev, A., \& Reshetnikov, V.\ 2013, \aap, 554, A11 
\bibitem[Conselice et al.(2003)]{Conselice2003} Conselice, C.~J., Bershady, M.~A., Dickinson, M., \& Papovich, C.\ 2003, \aj, 126, 1183 
\bibitem[De Lucia et al.(2006)]{Delucia_2006} De Lucia, G., Springel, V., White, S.~D.~M., Croton, D., \& Kauffmann, G.\ 2006, \mnras, 366, 499 
\bibitem[di Serego Alighieri et al.(2005)]{Diserego_2005} di Serego Alighieri, S., Vernet, J., Cimatti, A., et al.\ 2005, \aap, 442, 125 
\bibitem[{{Faber}(1973)}]{Faber_1973}
{Faber} S.~M., 1973, \apj, 179, 731
\bibitem[{{Faber} {et~al}\mbox{.}(2007){Faber}, {Willmer}, {Wolf}, {Koo},
  {Weiner}, {Newman}, {Im}, {Coil}, {Conroy}, {Cooper}, {Davis}, {Finkbeiner},
  {Gerke}, {Gebhardt}, {Groth}, {Guhathakurta}, {Harker}, {Kaiser}, {Kassin},
  {Kleinheinrich}, {Konidaris}, {Kron}, {Lin}, {Luppino}, {Madgwick},
  {Meisenheimer}, {Noeske}, {Phillips}, {Sarajedini}, {Schiavon}, {Simard},
  {Szalay}, {Vogt}, \& {Yan}}]{Faber_2007}
{Faber} S.~M. {et~al.}, 2007, \apj, 665, 265
\bibitem[Feldmann et al.(2008)]{Feldmann_2008} Feldmann, R., Mayer, L., \& Carollo, C.~M.\ 2008, \apj, 684, 1062-1074 
\bibitem[Galletta et al.(1997)]{Galletta_1997} Galletta, G., Sage, L.~J., \& Sparke, L.~S.\ 1997, \mnras, 284, 773 
\bibitem[George \& Zingade(2015)]{George15} George, K., \& Zingade, K.\ 2015, \aap, 583, A103 
\bibitem[Huertas-Company et al.(2010)]{Huertas_2010} Huertas-Company, M., Aguerri, J.~A.~L., Tresse, L., et al.\ 2010, \aap, 515, A3 
\bibitem[Iodice et al.(2002a)]{Iodice_2002a} Iodice, E., Arnaboldi, M., De Lucia, G., et al.\ 2002, \aj, 123, 195 
\bibitem[Iodice et al.(2002b)]{Iodice_2002b} Iodice, E., Arnaboldi, M., Sparke, L.~S., Gallagher, J.~S., \& Freeman, K.~C.\ 2002, \aap, 391, 103 
\bibitem[Iodice et al.(2002c)]{Iodice_2002c} Iodice, E., Arnaboldi, M., Sparke, L.~S., \& Freeman, K.~C.\ 2002, \aap, 391, 117 
\bibitem[Kannappan et al.(2009)]{Kannappan_2009} Kannappan, S.~J., Guie, J.~M., \& Baker, A.~J.\ 2009, \aj, 138, 579 
\bibitem[Kaviraj et al.(2009)]{Kaviraj_2009} Kaviraj, S., Peirani, S., Khochfar, S., Silk, J., \& Kay, S.\ 2009, \mnras, 394, 1713 
\bibitem[Kaviraj(2010)]{Kaviraj_2010} Kaviraj, S.\ 2010, \mnras, 408, 170 
\bibitem[Kaviraj et al.(2011)]{Kaviraj_2011} Kaviraj, S., Tan, 
K.-M., Ellis, R.~S., \& Silk, J.\ 2011, \mnras, 411, 2148 
\bibitem[Kaviraj(2014)]{Kaviraj_2014} Kaviraj, S.\ 2014, \mnras, 
437, L41 
\bibitem[Kennicutt(1998)]{Kennicutt_1998} Kennicutt, R.~C., Jr.\ 1998, \apj, 498, 541 
\bibitem[Kim et al.(2012)]{Kim_2012} Kim, T., Sheth, K., Hinz, J.~L., et al.\ 2012, \apj, 753, 43 
\bibitem[{{Komatsu} {et~al}\mbox{.}(2011){Komatsu}, {Smith}, {Dunkley},
  {Bennett}, {Gold}, {Hinshaw}, {Jarosik}, {Larson}, {Nolta}, {Page},
  {Spergel}, {Halpern}, {Hill}, {Kogut}, {Limon}, {Meyer}, {Odegard}, {Tucker},
  {Weiland}, {Wollack}, \& {Wright}}]{Komatsu_2011}
{Komatsu} E. {et~al.}, 2011, \apjs, 192, 18
\bibitem[Kormendy(1977)]{Kormendy_1977} Kormendy, J.\ 1977, \apj, 218, 333 
\bibitem[Willett et al.(2013)]{Willett_2013} Willett, K.~W., Lintott, C.~J., Bamford, S.~P., et al.\ 2013, \mnras, 435, 2835 
\bibitem[La Barbera et al.(2003)]{Labarbera_2003} La Barbera, F., Busarello, G., Merluzzi, P., Massarotti, M., \& Capaccioli, M.\ 2003, \apj, 595, 127 
\bibitem[Lintott et al.(2008)]{Lintott_2008} Lintott, C.~J., Schawinski, K., Slosar, A., et al.\ 2008, \mnras, 389, 1179 
\bibitem[Mapelli et al.(2015)]{Mapelli_2015} Mapelli, M., Rampazzo, R., \& Marino, A.\ 2015, \aap, 575, A16 
\bibitem[Marino et al.(2011)]{Marino_2011} Marino, A., Bianchi, L., Rampazzo, R., et al.\ 2011, \apj, 736, 154 
\bibitem[McIntosh et al.(2014)]{McIntosh14} McIntosh, D.~H., Wagner, C., Cooper, A., et al.\ 2014, \mnras, 442, 533 
\bibitem[Meert et al.(2013)]{meert_2013} Meert, A., Vikram, V., \& Bernardi, M.\ 2013, \mnras, 433, 1344 
\bibitem[Meert et al.(2015)]{meert_2015} Meert, A., Vikram, V., \& Bernardi, M.\ 2015, \mnras, 446, 3943 
\bibitem[Morganti et al.(2006)]{Morganti2006} Morganti, R., de Zeeuw, P.~T., Oosterloo, T.~A., et al.\ 2006, \mnras, 371, 157 
\bibitem[Moffett et al.(2012)]{Moffett_2012} Moffett, A.~J., Kannappan, S.~J., Baker, A.~J., \& Laine, S.\ 2012, \apj, 745, 34 
\bibitem[Moffett et al.(2015)]{Moffett_2015} Moffett, A.~J., Kannappan, S.~J., Berlind, A.~A., et al.\ 2015, \apj, 812, 89 
\bibitem[Nair \& Abraham(2010)]{Nair_2010} Nair, P.~B., \& Abraham, R.~G.\ 2010, \apjs, 186, 427
\bibitem[Oosterloo et al.(2010)]{Oosterloo2010} Oosterloo, T., Morganti, R., Crocker, A., et al.\ 2010, \mnras, 409, 500 
\bibitem[{{Peng} {et~al}\mbox{.}(2002){Peng}, {Ho}, {Impey}, \&
  {Rix}}]{Peng_2002}{Peng} C.~Y., {Ho} L.~C., {Impey} C.~D., {Rix} H.-W., 2002, \aj, 124, 266
\bibitem[Peirani et al.(2010)]{Peirani_2010} Peirani, S., Crockett, R.~M., Geen, S., et al.\ 2010, \mnras, 405, 2327 
\bibitem[Quinn(1984)]{Quinn_1984} Quinn, P.~J.\ 1984, \apj, 279, 596 
\bibitem[Reda et al.(2004)]{Reda_2004} Reda, F.~M., Forbes, D.~A., Beasley, M.~A., O'Sullivan, E.~J., \& Goudfrooij, P.\ 2004, \mnras, 354, 851 
\bibitem[Rutkowski et al.(2014)]{Rutkowski_2014} Rutkowski, M.~J., Jeong, H., Cohen, S.~H., et al.\ 2014, \apj, 796, 101 
\bibitem[Schawinski et al.(2009)]{schawinski09} Schawinski, K., 
Lintott, C., Thomas, D., et al.\ 2009, \mnras, 396, 818 
\bibitem[Serra \& Oosterloo(2010)]{Serra_2010} Serra, P., \& Oosterloo, T.~A.\ 2010, \mnras, 401, L29 
\bibitem[Serra et al.(2014)]{Serra2014} Serra, P., Oser, L., Krajnovi{\'c}, D., et al.\ 2014, \mnras, 444, 3388 
\bibitem[Sersic(1968)]{Sersic1968} Sersic, J.~L.\ 1968, Cordoba, Argentina: Observatorio Astronomico, 1968
\bibitem[Stark et al.(2013)]{Stark_2013} Stark, D.~V., Kannappan, S.~J., Wei, L.~H., et al.\ 2013, \apj, 769, 82 
\bibitem[Thilker et al.(2010)]{Thilker_2010} Thilker, D.~A., Bianchi, L., Schiminovich, D., et al.\ 2010, \apjl, 714, L171 
\bibitem[Toomre(1977)]{Toomre_77} Toomre, A.\ 1977, Evolution of Galaxies and Stellar Populations, 401 
\bibitem[Trager et al.(2000)]{Trager_2000} Trager, S.~C., Faber, 
S.~M., Worthey, G., \& Gonz{\'a}lez, J.~J.\ 2000, \aj, 120, 165 
\bibitem[Trujillo et al.(2011)]{Trujillo_2011} Trujillo, I., Ferreras, I., \& de La Rosa, I.~G.\ 2011, \mnras, 415, 3903 
\bibitem[van Driel et al.(2000)]{vanDriel_2000} van Driel, W., Arnaboldi, M., Combes, F., \& Sparke, L.~S.\ 2000, \aaps, 141, 385 
\bibitem[van Driel et al.(2002)]{vanDriel_2002} van Driel, W., Combes, F., Arnaboldi, M., \& Sparke, L.~S.\ 2002, \aap, 386, 140 
\bibitem[Vikram et al.(2010)]{Vikram_2010} Vikram, V., Wadadekar, Y., Kembhavi, A.~K., \& Vijayagovindan, G.~V.\ 2010, \mnras, 409, 1379 
\bibitem[Visvanathan 
\& Sandage(1977)]{Visvanathan_77} Visvanathan, N., \& Sandage, A.\ 1977, \apj, 216, 214 
\bibitem[Wei et al.(2010)]{Wei_2010} Wei, L.~H., Vogel, S.~N., Kannappan, S.~J., et al.\ 2010, \apjl, 725, L62 
\bibitem[Wong et al.(2015)]{Wong_2015} Wong, O.~I., Schawinski, K., J{\'o}zsa, G.~I.~G., et al.\ 2015, \mnras, 447, 3311 
\bibitem[Whitmore et al.(1990)]{Whitmore_1990} Whitmore, B.~C., Lucas, R.~A., McElroy, D.~B., et al.\ 1990, \aj, 100, 1489 
\bibitem[Yi et al.(2005)]{Yi_2005} Yi, S.~K., Yoon, S.-J., Kaviraj, S., et al.\ 2005, \apjl, 619, L111 
\bibitem[Young et al.(2011)]{Young_2011} Young, L.~M., Bureau, M., 
Davis, T.~A., et al.\ 2011, \mnras, 414, 940 
\bibitem[Young et al.(2014)]{Young_2014} Young, L.~M., Scott, N., 
Serra, P., et al.\ 2014, \mnras, 444, 3408 

\end{thebibliography}
\end{document}